\documentclass[a4paper,fleqn]{cas-dc}

\usepackage[numbers]{natbib}
\usepackage{natbib}
\usepackage{amsmath,amssymb,amsfonts}
\usepackage{algorithmic}
\usepackage{graphicx}
\usepackage{textcomp}
\usepackage{bigstrut}
\usepackage{multirow}
\usepackage{array}
\usepackage{todonotes}
\usepackage{comment}
\usepackage{acronym}

\def\tsc#1{\csdef{#1}{\textsc{\lowercase{#1}}\xspace}}
\tsc{WGM}
\tsc{QE}
\tsc{EP}
\tsc{PMS}
\tsc{BEC}
\tsc{DE}

\acrodef{lvh}[LVH]{left ventricular hypertrophy}
\acrodef{rvh}[RVH]{right ventricular hypertrophy}
\acrodef{lae}[LAE]{left atrial enlargement}
\acrodef{std}[STD]{ST depression}
\acrodef{std}[STE]{ST elevation}
\acrodef{ecg}[ECG]{electrocardiogram}
\acrodef{snr}[SNR]{signal-to-noise ratio}
\acrodef{mse}[MSE]{mean squared error}
\acrodef{stderr}[STDERR]{standard error}
\acrodef{lbbb}[LBBB]{left bundle branch block}
\acrodef{rbbb}[RBBB]{right bundle branch block}

\begin{document}
\let\WriteBookmarks\relax
\def\floatpagepagefraction{1}
\def\textpagefraction{.001}

\shorttitle{Diagnostic Quality Assessment}

\shortauthors{Peter Kovács et~al.}

\title [mode = title]{Diagnostic Quality Assessment for Low-Dimensional ECG Representations}                      

\nonumnote{This work was supported by the Upper Austrian Medical Cognitive Computing Center (MC$^3$) and by the \'UNKP-21-5 New National Excellence Program of the Ministry for Innovation and Technology of Hungary from the source of the National Research, Development and Innovation Fund and by the University SAL Labs initiative of Silicon Austria Labs (SAL).} 


%
\author[1]{Péter Kovács}

\cormark[1]

\fnmark[1]

\ead{kovika@inf.elte.hu}



\affiliation[1]{organization={Department of Numerical Analysis, E\"otv\"os Lor\'and University},
    addressline={P\'azm\'any P\'eter s\'et\'any 1/c.}, 
    city={Budapest},
    postcode={1117}, 
    country={Hungary}}

\author[2]{Carl B\"ock}[]
\fnmark[1]
\ead{carl.boeck@jku.at}

\affiliation[2]{organization={JKU LIT SAL eSPML Lab, Institute of Signal Processing, Johannes Kepler University Linz},
    addressline={Altenberger Strasse 69}, 
    city={Linz},
    postcode={4040}, 
    country={Austria}}

\author[3]{Thomas Tschoellitsch}[]
\ead{Thomas.Tschoellitsch@kepleruniklinkum.at}

\affiliation[3]{organization={Clinic of Anesthesiology and Intensive Care Medicine, Johannes Kepler University Linz},
    addressline={Krankenhausstraße 9}, 
    city={Linz},
    postcode={4020}, 
    country={Austria}}



\author[4]{Mario Huemer}
\ead{mario.huemer@jku.at}

\affiliation[4]{organization={Institute of Signal Processing, Johannes Kepler University Linz},
    addressline={Altenberger Straße 69}, 
    city={Linz},
    postcode={4040}, 
    country={Austria}}

\author[3]{Jens Meier}[]
\ead{Jens.Meier@kepleruniklinkum.at}

\cortext[cor1]{Corresponding author}

\fntext[fn1]{These authors contributed equally to this work.}


\begin{abstract}
There have been several attempts to quantify the diagnostic distortion caused by algorithms that perform low-dimensional electrocardiogram (ECG) representation. However, there is no universally accepted quantitative measure that allows the diagnostic distortion arising from denoising, compression, and ECG beat representation algorithms to be determined. Hence, the main objective of this work was to develop a framework to enable biomedical engineers to efficiently and reliably assess diagnostic distortion resulting from ECG processing algorithms. We propose a semiautomatic framework for quantifying the diagnostic resemblance between original and denoised/reconstructed ECGs. Evaluation of the ECG must be done manually, but is kept simple and does not require medical training. In a case study, we quantified the agreement between raw and reconstructed (denoised) ECG recordings by means of kappa-based statistical tests. The proposed methodology takes into account that the observers may agree by chance alone. Consequently, for the case study, our statistical analysis reports the "true", beyond-chance agreement in contrast to other, less robust measures, such as simple percent agreement calculations. Our framework allows efficient assessment of clinically important diagnostic distortion, a potential side effect of ECG (pre-)processing algorithms. Accurate quantification of a possible diagnostic loss is critical to any subsequent ECG signal analysis, for instance, the detection of ischemic ST episodes in long-term ECG recordings.
\end{abstract}


\maketitle

\section{Introduction}
\label{sec:introduction}

In medical applications, expert systems require both the extraction of reliable information from biomedical signals and efficient representation of this knowledge. Extracting relevant features is essential, since these are usually either fed to rule-based expert systems or machine learning methods or used directly by medical experts to derive a diagnosis. Clearly, the main objective of biomedical signal processing methods is therefore to efficiently represent important medical knowledge, eliminating noisy and redundant signal features while keeping the informative ones. However, applying denoising or feature reduction algorithms may lead to unintentional removal of diagnostic information. Common metrics in signal processing, such as \ac{snr}, \ac{mse}, and \ac{stderr}, quantify the numerical error, but not the loss of diagnostic information. For instance, in the case of evaluating an electrocardiogram (ECG) that is superimposed with noise (e.g., power-line interference or baseline wander), the values of the previously mentioned objective measures may improve significantly if denoising algorithms are applied or the signal is represented in a low-dimensional space. However, these measures do not consider possible diagnostic distortions that may change the interpretation of the \ac{ecg} curves, and -- as a consequence -- the diagnosis. Signal-quality indices (SQI) form another class of metrics which lies in between quantitative and qualitative distortion measures. These application-oriented metrics are intended to quantify the suitability of ECG signals for deriving reliable estimation of particular medical features, such as the heart rate~\cite{qualass1, qualass2}. Note that the use cases of SQIs are limited to the application in question, and thus they are not applicable to perform a thorough agreement analysis between diagnostic features of the raw ECG and its low-dimensional representations. In fact, to date there is no universally accepted quantitative measure that captures the diagnostic distortion of such biomedical signal processing methodologies.

In this paper, we propose a testing framework that is suitable for measuring the diagnostic resemblance between preprocessed ECG signals and original recordings. This enables medical validation of ECG processing algorithms, such as filtering and data compression, and extraction/monitoring of clinical features. The latter has recently grown in importance, mainly due to the increasing amounts of data recordable (e.g., long-term ECG recordings), which require automated information extraction in order to be manageable by medical experts. In addition to standard clinical features (e.g., QT interval, QRS duration), the shapes of individual waves and characteristic shape changes over time are of high diagnostic interest and have been shown to carry important information \cite{Laguna2016}. However, it is exactly this information that might get lost when applying algorithms for low-dimensional information representation. Eliminating noisy signal features, such as the baseline wander, may lead inadvertently to removal of ischemic ST episodes \cite{Lenis2017}, which has a direct influence on diagnosis. Further, wave-shape feature interpretation might change: for instance, an originally positive wave may then be identified as a biphasic one. Consequently, a methodology is needed that enables quantitative representation of such potentially diagnostically relevant changes. 
However, reliable evaluation of whether diagnostic information has changed due to algorithmic processing will always require human expertise. Our test design minimizes the workload and does not require the evaluating person to be a medically qualified, since the test involves deciding between distinct wave shape types rather than making a specific diagnosis. Biomedical engineers can therefore easily carry out this evaluation themselves and can improve their algorithms more rapidly and efficiently, as they do not have to wait for feedback from medical experts who are able to diagnose a possibly very specific pathology. We demonstrate this by comparing original ECG recordings (selected from \cite{Welch1991}) and their corresponding low-dimensional representations, which were obtained by applying an approach we have previously developed \cite{carlpeter}. 
Our framework is intended to help biomedical engineers in evaluating the diagnostic distortion of an ECG processing algorithm for real-world recordings in the early stages of development. In contrast to \cite{carlpeter}, where we evaluated diagnostic distortion based on synthetic data only, in this work we elaborate on real \ac{ecg} recordings, as they appear in daily clinical practice. Although synthetic data is crucial in biomedical signal processing, since they provide the ground truth (clean signal), these signals do not capture the wide variety of possible \ac{ecg} morphologies and noise. It should be emphasized, however, that alongside the proposed diagnostic distortion analysis, a complete performance report must asses the robustness of the tested algorithms with respect to the noise level~\cite{cinc84}, the perturbation of the input parameters~\cite{stdep}, etc.

The methodology we propose for agreement analysis is inspired by former work on diagnostic distortion measures, which we review in Section~\ref{sec:relwork}. The design concepts of our testing framework are discussed in Section~\ref{sec:testdesign}. Section~\ref{sec:experiments} briefly describes our previous work on low-dimensional ECG representations~\cite{tbme_paper}, followed by an analysis of the agreement between the diagnostic features of original ECGs and their approximations. Finally, Sections~\ref{sec:discussions} and \ref{sec:conc} discuss the results and conclude the paper, respectively.

\section{Related work \label{sec:relwork}}
The approaches most closely related to ours seek to quantify diagnostic relevance in ECG signal compression. Compression is very similar to feature extraction in that ECG data samples are represented in a low-dimensional (feature) space from which the original ECG signal can be restored. In order to prove that the reconstruction preserves diagnostic information, the quality of the restored ECG signal must be evaluated. 

\begin{figure}[!t]
\includegraphics[width=\columnwidth]{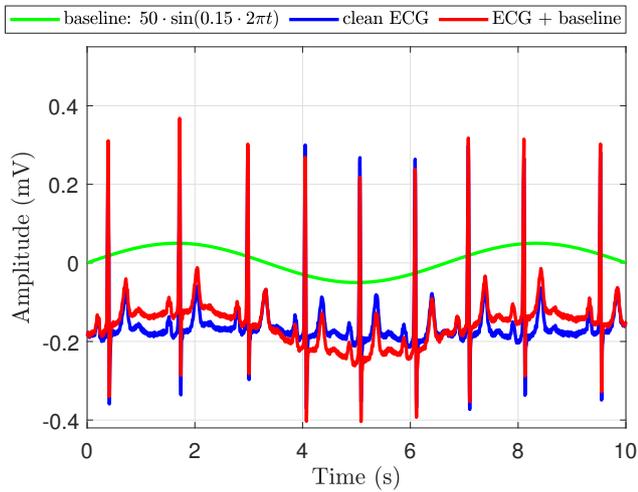}
\caption{Analyzing the diagnostic distortion after perfect baseline removal, using objective measures: PRD=$20.4\%$, WWPRD=$31.2\%$, WEDD=$26.0\%$, which corresponds to the quality groups \textit{not bad} and \textit{bad} according to Tab.~8 in \cite{wedd}.}
\label{fig:objmetrics}
\end{figure}

Diagnostic distortion can be measured by objective methods that are based on mathematical models. The most commonly used objective evaluation method uses the percent root-mean-square difference (PRD), measuring the squared error between original and reconstructed ECG signal. The PRD is a numerical quantity that assumes equal error contribution over the whole ECG. This is not appropriate for evaluating diagnostic distortion, since numerical error of the same degree, for instance, in the approximations of the QRS complex and the P wave do not imply the same level of diagnostic error. The weighted diagnostic distortion (WDD) measure was an early attempt to tackle this problem and to quantify the diagnostic error of compressed ECG signals \cite{wdd}. It compares amplitude, duration, and shape features of original and reconstructed ECGs, and assigns weights to the corresponding error terms based on their diagnostic relevance. These features are, however, extracted automatically, and thus the procedure is prone to inaccuracies of the ECG delineation algorithms. Later, Al-Fahoum \cite{wwprd} introduced the wavelet-based weighted PRD (WWPRD), which quantifies the diagnostic error in wavelet space: In each wavelet subband, the PRD between the wavelet coefficients of the original and the reconstructed ECG is calculated, and then the error contribution of each subband is weighted based on its diagnostic significance. Although the WWPRD can be easily calculated and correlates very well with clinically evaluated results, it is heavily influenced by the presence of noise in the relevant subbands. Manikandan et al.~\cite{wedd} therefore proposed the wavelet-energy-based diagnostic distortion (WEDD) measure, a reweighted variant of the WWPRD. In WEDD, the weights are equal to the relative energy between the overall signal and the corresponding subbands. This way low-energy high-frequency noise can be suppressed in the PRD calculation. However, low-frequency high-energy noise which overlaps with the diagnostic content of the ECG is counted in the WEDD. Fig.~\ref{fig:objmetrics} illustrates this phenomena for an ECG distorted by a sinusoidal baseline drift. Even though the diagnostic remained unchanged, the WEDD, WWPRD, and PRD measures suggest a significant distortion of the ECG curve. All these metrics have their limitations in medical validation, as discussed in relation to baseline removal algorithms. 

Clearly, application of such objective evaluation methods does not require the feedback from cardiologists, and thus the test results are not influenced by intra- and interobserver variability. Despite their advantages, objective distortion measures are not fully accepted in the ECG signal processing community due to their lack of medical validation \cite{ECGmetrics_review}. Subjective methods, such as the mean opinion score (MOS), are therefore considered to be the gold standard \cite{ECGmetrics_review}. In this case, quality is assessed by cardiologists, who check whether original and reconstructed signals imply the same diagnosis. The results of these tests are then converted into a single measure called the MOS error value, which quantifies the diagnostic distortion of the compression/feature extraction technique under study. The MOS test introduced by Zigel et al.~\cite{wdd} and its variant \cite{mostest} remain in use for evaluating ECG processing algorithms. In these studies, cardiologists rated the general quality of the ECG recordings, and investigated whether the same interpretation would result from the medical features of the original and the processed ECGs. To this end, a set of ECG features was considered which included both simple shape features, such as positivity/negativity of the T wave, and more complex morphological characteristics, such as left/right bundle branch block (LBBB/RBBB) and premature ventricular contraction (PVC). However, some of these features are ambiguous (e.g., waveform symmetry, which is difficult to distinguish from slightly asymmetric cases), and thus increase intra- and interobserver variability. Other morphological characteristics are just too complex to identify without the help of clinicians. Another drawback of previous MOS tests is that they consider only the overall percentage of agreement and ignore the possibility that the observers may agree by chance alone. For instance, the presence of a delta wave in the QRS complex is relatively rare \cite{Soria1982}, and therefore in most cases cardiologists will simply exclude this diagnosis.        

We developed a carefully designed MOS test for evaluating the diagnostic relevance of ECG preprocessing algorithms that uses only simple shape features which can be recognized by biomedical experts and which reduce the intra- and interobserver variability. The results of our test support design decisions and speed up the development of ECG processing algorithms, since input from the medical experts is then only needed in the final testing phase. Note that current approaches consider the proportion of observed agreement alone as an index of concordance. we, in contrast suggest the use of Cohen's kappa in assessing the performance of ECG preprocessing algorithms, as it takes into account agreement by chance \cite{kappa_studies1, kappa_studies2, kappa_studies3}.




\section{Test design \label{sec:testdesign}}
The main objective of the test design was to allow fast and reliable evaluation of the diagnostic distortion caused by low-dimensional ECG signal representation. Together with a medical expert we selected 26 recordings from the Massachusetts general hospital/Marquette fundation (MGHMF) database \cite{Welch1991, PhysioNet}. This database has a detailed patient guide, which allows recordings to be chosen according to the occurrence of various pathologies and waveforms. An overview of the selected recordings is given in Table~\ref{tab:ph_recs}; for more detailed information we refer to the patient guide provided in \cite{PhysioNet}. Our main inclusion criterion for selecting the recordings was the occurrence of various (abnormal) wave morphologies, for instance, positive/negative, biphasic, and flattened waves. Further, different manifestations of the QRS complex (R, Rs, RS, etc.) and the ST segment (e.g., elevated or depressed) were important criteria, since these are key diagnostic features which should be preserved by ECG compression / denoising algorithms. A medical expert chose the recordings based on the patient guide and visual judgement of the \ac{ecg} strips.

\begin{table}
\setlength{\tabcolsep}{3pt}
    \centering
    \begin{scriptsize}
    \caption{Recordings selected for evaluating the ECG beat representation algorithm \cite{Welch1991}.}

    \begin{tabular}{|c|c|c|l|}
    \hline
         \textbf{recordname} & \textbf{age} & \textbf{sex} & \textbf{main ECG morphological interpretation}\\ \hline
         mgh003 & 47 & f & poor R wave progression, \\ 
         & & & ST/T abnormalities \\ \hline
         mgh007 & 60 & f & normal, strong baseline wander \\ \hline
         mgh016 & 82 & m & ST/T abnormalities, possible old infarction\\ \hline
         mgh021 & 75 & m & possible acute anterior subendocardial \\ & & & injury and possibly evolving infarction \\ \hline
         mgh034 & 66 & m & old inferior infarct, left atrial enlargement \\ \hline
         mgh037 &  71 & f & \ac{lvh}, \\
         & & & old anterior infarct \\ \hline
         mgh043 & 66 & m & acute ST wave changes, inferior wall infarct \\ \hline
         mgh056 &  70 & m & prolonged QT, old interior infarct \\ \hline
         mgh089 & 32 & f & sinus tachycardia \\ \hline
         mgh095 & 16 & m &  ST / T wave changes \\ \hline
         mgh097 & 63 & m & \ac{lae}, old infarct \\ \hline
         mgh099 &  49 & m & old infarct, ST/T wave changes\\ \hline
         mgh100 & 20 & m & sinus tachykardia \\ \hline
         mgh113 & 70 & m &  old infarct, ST/T wave changes \\ \hline
         mgh134 & 71 & m & left anterior hemiblock, left axis deviation \\ \hline
         mgh137 & 60 & f & low voltage, ST/T wave abnormalities \\ \hline
         mgh156 & 41 & m & right ventricular strain, \ac{lae} \\ \hline
         mgh164 &  62 & f & \ac{rvh} or \\
         & & &left posterior hemiblock \\ \hline
         mgh174 & 27 & m & \ac{lvh}, \ac{lae} \\ \hline
         mgh184 & 72 & m &   normal \\ \hline
         mgh193 & 62 & f & subendocardial ischemia, digitalis effect\\ \hline
         mgh195 & 29 & m & slight J point elevation \\ \hline
         mgh203 &  41 & f & postoperative ST/T wave changes\\ \hline
         mgh209 & 62 & m & rhythm disturbances \\ \hline
         mgh247 & 73 & f & old interior infarct, ST/T wave changes \\ \hline
         mgh250 & - & - &\ac{std} \\ \hline
    \end{tabular}
    \label{tab:ph_recs}
    \end{scriptsize}
\end{table}

\begin{figure*}[htb]
\centering
\hspace*{-2.9cm}
\includegraphics[width=2.7\columnwidth]{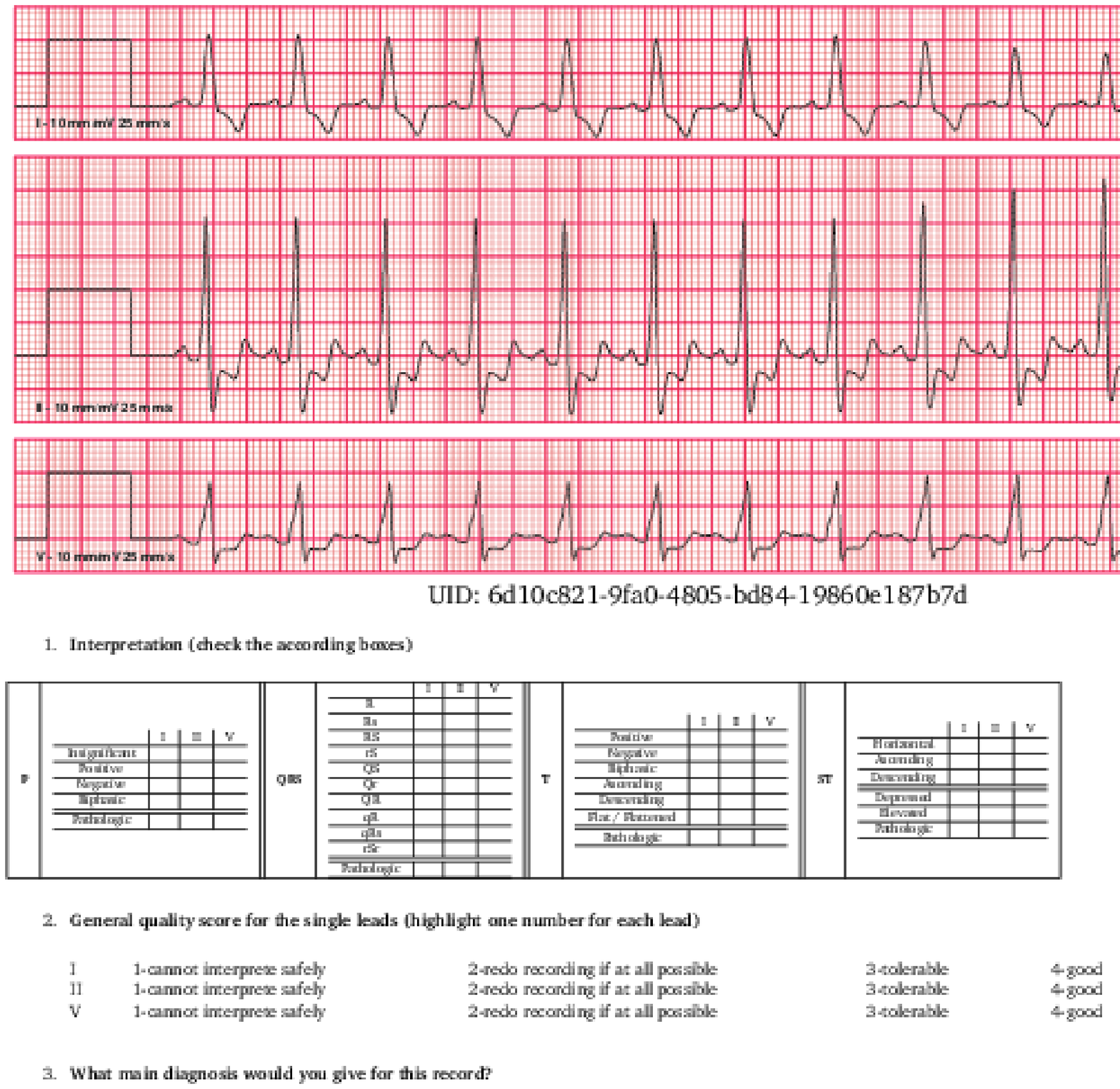}
\vspace*{-2cm}
\caption{Example recording for the evaluation of ECG preprocessing / compression algorithms. The single leads are investigated according to the wave shapes and possible pathologies present, before judging the general quality of the ECG lead and optionally giving a main diagnosis. }
\label{fig:mos_test_example}
\end{figure*}

Subsequently, as described in Sec.~\ref{sec:formerwork}, these recordings were transformed into a low-dimensional representation using Hermite and sigmoidal functions combined with spline interpolation \cite{carlpeter}. These functions have been extensively studied in several ECG related medical applications, such as data compression~\cite{hexp5}, heartbeat clustering~\cite{hexp4}, and myocardial infarction detection~\cite{hexp1}. In order to investigate the diagnostic distortion of Hermite-based ECG decomposition, a test set was built that included 32 original 3-lead ECG recordings and 32 reconstructed 3-lead ECG recordings, 12 of which (6 original and 6 reconstructed) occured twice. This was done to allow assessment of self-consistency (within-observer agreement). In total, 64 3-lead ECG recordings (i.e., 192 ECG strips), were therefore evaluated by experts. The test set was split into 4 subsets, each of which was to be processed on a different day to avoid exhaustion and possible resulting inaccuracies that would bias the results. The recordings were arranged in a pseudo-randomised order with the restriction that original and reconstructed ECG recordings were not allowed to occur in the same subset. To simulate daily clinical practice, the ECG recordings were presented on a standard ECG grid,(10~mm/mV and 25~mm/s). An example recording is shown in Fig.~\ref{fig:mos_test_example} (note that this recording is scaled for better visibility). The questionnaire for assessing diagnostic distortion was then designed based mainly on two factors:


First, in our preliminary experiments, we realized that even highly experienced physicians were not able to identify with confidence more complex pathologies such as a \ac{lbbb} or a \ac{rbbb} based on the ECG recordings alone. This is because additional laboratory tests or additional ECG leads would be needed for a sufficiently accurate diagnosis. Therefore, if the questionnaire offers options such as \ac{lbbb} and \ac{rbbb}, experts tend not to tick these boxes unless it is a very clear case, which could of course bias the evaluation significantly. The results may lead one to believe that the reconstructed (low-dimensional) ECG still  has retained the complete diagnostic information, but this might just be due to the ECG always being labeled as normal by the expert. 

Second, development of a signal representation algorithm should require medical expertise only in the final testing phase. We therefore included only simple evaluation criteria for judging the diagnostic distortion of the P and T waves, the QRS complex, and the ST segment, as illustrated in Fig.~\ref{fig:mos_test_example}.

Specifically, this means that in a first step the ECG wave segments of all available leads are to be evaluated according to their general shape features, for instance, insignificant, positive, negative, and biphasic in the case of the P wave. Clearly, one of these options must be selected. Additionally, depending on the segment investigated, the evaluating person may tick an optional box which indicates a (general) pathology indicated by the wave. 

Subsequently, the quality of the single leads is to be rated, where the experts are asked to focus on the signal clarity of the wave. This allows assessment of whether the low-dimensional representation degrades, retains, or even increases the quality of the ECG recording. Quality improvement would imply that noisy signal features were successfully eliminated while important diagnostic features were retained. Finally, a main diagnosis is to be given as free text. However, it should be mentioned, that this is considered optional and should only be answered by medical experts (in the final testing stage). This allows assessing whether the main diagnosis changed between ECG recording and low-dimensional signal representation and serves as an additional source for identifying a possible diagnostic distortion. 

For our case study a total of 3 physicians were briefed with the information above and with additional instructions in written (see supplementary material) and oral form. 

\section{Case-study: low-dimensional ECG representation \label{sec:experiments}}
In the best case, low-dimensional ECG representation preserves important diagnostic/morphological features, while redundant and noisy signal features are mostly eliminated. This is, however, a difficult task, since the frequency spectra of the ECG and possible noise (e.g., baseline wander) overlap in most cases \cite{sornmobook}. Therefore, the diagnostic distortion of these preprocessing techniques must be investigated before they are applied to real-world problems. In this study, we tested the reliability of our former work approach (Sec.~\ref{sec:formerwork}) by assessing its between-method agreement. More specifically, by means of statistical tests we checked whether two measurements (i.e., the original, and the reconstructed low-dimensional signal) produce the same diagnostic features defined in our MOS test.   

In order to estimate the consensus between original and reconstructed ECGs, we computed the proportion of observed agreement $P_o$ and the $\kappa$ coefficients between each (original and reconstructed) feature pair. In the case of dichotomous features, these quantities can be calculated as follows:
\begin{equation}
    \kappa=\frac{P_o-P_c}{1-P_o}, \quad P_o=\frac{a+d}{n}, \quad P_c=\frac{f_1 g_1+f_2g_2}{n^2},
    \label{eq:kappa}
\end{equation}
where $P_c$ denotes the chance agreement, $f_i, g_i$ are the marginal totals, and $a$ and $d$ are the numbers of agreements on \textit{present} and \textit{absent} values of the corresponding feature (see Tab.~\ref{tab:binary_contab}). Note that, in the case of morphological features, we defined more than two mutually exclusive categories for which the calculations in Eq.~\eqref{eq:kappa} can be generalized according to \cite{genkappa1, genkappa2}. Cohen's kappa is widely used in reliability studies in clinical research \cite{kappa_studies1, kappa_studies2, kappa_studies3}. The kappa values range from $-1$ to $1$; $\kappa=0$ suggests that the observed agreement is not better than would be expected by chance alone, while $\kappa=1$ implies perfect agreement, and negative values indicate potential systematic disagreement between the observers. Other values of kappa can be interpreted based on Tab.~\ref{tab:kappa_interpret} as proposed by Landis and Koch~\cite{kappa_interpret}.

In our case, achieving a perfect agreement (i.e.,~$\kappa=1$) is unrealistic because the evaluating cardiologists typically have different levels of experience and mental fatigue. Taking this into account, we also report $\kappa_{\max}$, which expresses the maximum attainable kappa provided that the marginal totals $f_i,g_i$ are fixed. The value of $\kappa_{\max}$ can easily be calculated by substituting $P_o$ with $P_{o,\max} = \big(\min (f_1,g_1) + \min (f_2,g_2)\big)/n$ in Eq.~\eqref{eq:kappa}. The difference $\kappa_{\max}-\kappa$ indicates the unachieved agreement
beyond chance constrained by the marginal
totals~\cite{kappa_studies1}.   

In ECG compression, it is common practice to provide $P_o$ as a measure of diagnostic concordance \cite{mostest}. However, $P_o$ results can be misleading, especially when most of the observations fall within a single category \cite{cons_est}. For instance, previous studies \cite{wdd, mostest} considered the presence of delta waves in the QRS complex; however, compared to other QRS shape features these occur relatively rarely \cite{Soria1982}. In most cases, cardiologists would therefore agree on the absence of delta waves in both the original and the filtered signals. To avoid such effects, alongside the percent-agreement figures, we report the corresponding $\kappa$ coefficients, their confidence intervals, and the maximum attainable kappa.   

\begingroup
\setlength{\tabcolsep}{3pt} 
\renewcommand{\arraystretch}{1.5} 
\begin{table}[t!]
\centering
\vspace{2mm}	
\caption{Contingency table in the case of dichotomous ECG features.}
\vspace{-2mm}
\begin{center}
\scalebox{0.89}{
	\begin{tabular}{|c|cc|c|} \hline
	\bigstrut
		\multirow{2}{6.5em}{\parbox{6.5em}{\vspace{0.3cm} \textbf{Feature of the original ECG}}} & \multicolumn{2}{|p{12em}|}{\centering\textbf{Feature of the filtered ECG}} & \multirow{2}{3.0em}{\parbox{3.0em}{\vspace{0.2cm} \textbf{Totals}}}\\
		\cline{2-3}
	\bigstrut
		& \multicolumn{1}{|p{6em}}{\centering \textbf{Present}} & \multicolumn{1}{|p{6em}|}{\centering \textbf{Absent}} & \\
		\cline{1-4}
 	    \bigstrut \textbf{Present} & \multicolumn{1}{|c|}{a} & b & $g_1$ \\ \cline{2-3}
		\bigstrut\textbf{Absent} & \multicolumn{1}{|c|}{c} & d & $g_2$ \\ \hline
		\bigstrut\textbf{Totals} & $f_1$ & $f_2$ & $n$ \\ \hline
	\end{tabular}
}
\end{center}
\label{tab:binary_contab}
\vspace{-5mm}	
\end{table}
\endgroup

\begingroup
\setlength{\tabcolsep}{3pt} 
\renewcommand{\arraystretch}{1.5} 
\begin{table}[t!]
\centering
\caption{Strength of agreement interpretation of kappa values \cite{kappa_interpret}.}
\scalebox{0.89}{
    \begin{tabular}{>{\centering\arraybackslash}p{0.2\textwidth}>{\centering\arraybackslash}p{0.2\textwidth}}
	\hline
        \centering \textbf{Kappa Statistic} & \textbf{Strength of Agreement}\\\hline
        \centering $<0.00$ & Poor \\
        \centering $0.00-0.20$ & Slight \\ 
        $0.21-0.40$ & Fair \\ 
        $0.41-0.60$ & Moderate \\ 
        $0.61-0.80$ & Substantial \\ 
        $0.81-1.00$ & Almost perfect\\ \hline
	\end{tabular}
}
\label{tab:kappa_interpret}
\vspace{-5mm}	
\end{table}
\endgroup

\subsection{\ac{ecg} beat representation - algorithm}
\label{sec:formerwork}
Among the classes of low-dimensional ECG representations, the Hermite-based decomposition has been widely studied especially for extracting features in advance to machine learning algorithms (see e.g., Chapter~12 in Ref.~\cite{ecgbook}). These approaches utilize similarities between the shapes of Hermite functions and ECG waveforms. In a recent work, we extended the theoretical framework of Hermite-based ECG models by sigmoidal functions combined with piecewise polynomial interpolation~\cite{carlpeter}. One of the main goals of this work was to reduce noisy and redundant signal features while retaining diagnostically important waveform features. Hence, we sought not only to reduce dimensionality, but also to simultaneously denoise the signal and to segment the \ac{ecg} into its fundamental waves (P-QRS-T). We used adaptive Hermite and sigmoidal functions to extract important \ac{ecg} waveform information, while piecewise polynomial interpolation captured mainly the undesired baseline wander. Additionally, because of the (smooth) basis functions we selected, high-frequency noise was also reduced. 

\begin{figure}[!ht]
\centerline{\includegraphics[width=\columnwidth]{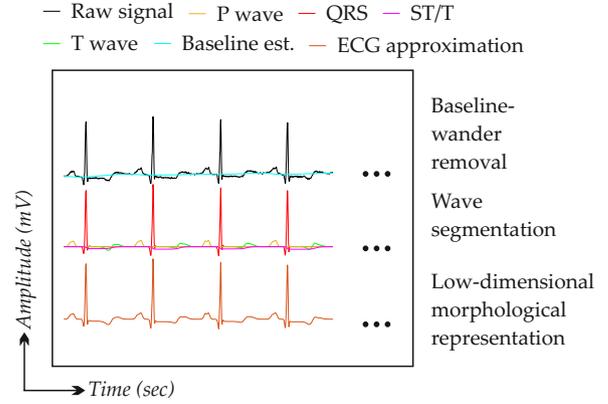}}
\caption{Simultaneous baseline wander removal, wave segmentation, and low-dimensional beat representation \cite{carlpeter}.}
\label{fig:former_work}
\end{figure}

Figure~\ref{fig:former_work} shows an example \ac{ecg} trace, which was segmented into its fundamental parts, that is, P wave, QRS complex, ST/T segment, T wave, and baseline estimation. As can be seen, the low-dimensional representation accurately describes the underlying \ac{ecg}.
This was achieved by developing a nonlinear least-squares model with an appropriate set of basis functions. This model was first tailored precisely to a single person by nonlinear global optimization, and then readjusted beat-by-beat by means of nonlinear local optimization. Optimization was carried out with respect to the translation and dilation of the basis functions used to represent the single waves (P-QRS-T). Thus, we created a person-specific model which allows tracking morphological changes in a low-dimensional space while retaining the characteristic shape information of the \ac{ecg} trace. 

\subsection{Experiments and results}
\label{sec:experiments_and_results}

Following the recommendations of Watson and Petrie \cite{kappa_studies2}, we conducted three experiments to evaluate the reliability of our previously published approach to low-dimensional \ac{ecg} representation \cite{carlpeter}. To this end, three cardiologists (C1, C2, C3), who were not allowed to discuss their answers with each other, filled out the blind test described in Section~\ref{sec:testdesign} and illustrated in Figure~\ref{fig:mos_test_example}. We partitioned the test records into four work packages to decrease the daily workload for the cardiologists and thus minimize factors that can negatively influence the evaluation, such as mental fatigue and lack of motivation. 
\subsubsection{Between-method agreement}
\begin{figure}[!t]
\centerline{\includegraphics[width=\columnwidth, trim=90 320 110 320, clip]{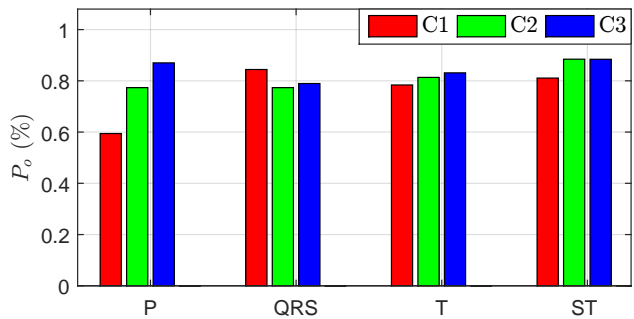}}
\caption{Between-method diagnostic concordance for the ECG features extracted from the original and the filtered signals.}
\label{fig:between_method_conc}
\end{figure}

\begin{figure}[!t]
\centerline{\includegraphics[width=\columnwidth, trim=90 320 110 320, clip]{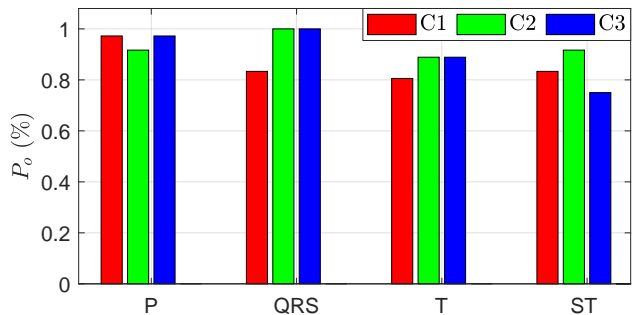}}
\caption{Between-method diagnostic concordance for the pathological waveshapes extracted from the original and the filtered signals.}
\label{fig:pathologic_conc}
\end{figure}

In this experiment, we studied the diagnostic concordance between the features of the original (raw) ECG and of the reconstructed signal \cite{carlpeter}.  Fig.~\ref{fig:between_method_conc} shows that the proportion of observed agreements $P_o$ is about $80\%$ for all cardiologists and all ECG features except for the P wave, where cardiologist C1 achieved only $60\%$ agreement. However, note that the self-consistency of C1 was also low for the P wave (see, e.g., Fig.~\ref{fig:within_observer_conc}). This is due to the P wave being a low-amplitude ECG component which has more ambiguous characteristics in the presence of noise than the QRS and the T waves.
This also explains why the between-method concordances in Fig.~\ref{fig:between_method_conc} vary considerably between cardiologists in the case of the P wave. We also evaluated the diagnostic concordance between the pathological wave shapes of the original and the reconstructed ECG signals. According to Fig.~\ref{fig:pathologic_conc}, the observed agreement was greater than $80\%$ in most cases, and close to $100\%$ for the P wave and for the QRS complex. Consequently, the low-dimensional beat representation investigated did not significantly increase ambiguity in terms of pathological versus non-pathological waveshape class.

We used kappa statistics to analyze the chance-corrected observed agreement between the features of the original and the filtered ECG (see Tab.~\ref{tab:between_method}). The highest $\kappa$ with the narrowest confidence intervals and the closest $\kappa_{\max}$ was achieved for the QRS complex. Second best was the T wave, with substantial agreement between the original ECG and the reconstructed signal (cf. Table~\ref{tab:kappa_interpret}). The kappa scores of the ST segment morphology are lower, which indicates fair agreement between the features. However, it seems that the low-dimensional representation preserved the ST depression and elevation, as indicated by high kappa scores. Furthermore, the maximum attainable kappa was observed for C2 and C3 in the case of ST depression. This corroborates our previous claims for our joint Hermite sigmoid model \cite{carlpeter}, namely, that the sigmoid functions perform well in modeling the on/offset shifts of the QRS complex and the ST elevation/depression. As with the diagnostic concordance in Fig.~\ref{fig:between_method_conc}, the kappa scores of the P wave morphology are not as consistent as the scores of the previously mentioned features. In this case, three different levels of agreement (i.e., fair, moderate, and substantial) can be observed between the original and the filtered ECG features. Note that none of the confidence intervals includes $0$, and we would thus reject the null hypothesis on $\kappa=0$. This means that no evidence of agreement by chance alone was found.


\begingroup
\setlength{\tabcolsep}{3pt} 
\renewcommand{\arraystretch}{1.5} 
\begin{table*}[t!]
\centering
\vspace{2mm}	
\caption{Using $\kappa$ statistics to analyze between-method agreement between original and filtered ECG features.}
\vspace{-2mm}
\begin{center}
\scalebox{0.73}{
	\begin{tabular}{|c|c|c|c|c|c|c||c|c|c|c|c|c|} \hline
	\bigstrut
		\multirow{2}{*}{\textbf{}} & \multicolumn{6}{|c||}{\textbf{P wave}} & \multicolumn{6}{|c|}{\textbf{QRS complex}}\\
		\cline{2-13}
	\bigstrut
		& \multicolumn{3}{|c|}{$\boldsymbol{\kappa}$ ($\boldsymbol{\kappa_{\max}})$} & \multicolumn{3}{|c||}{\textbf{Confidence intervals}} & 		\multicolumn{3}{|c|}{$\boldsymbol{\kappa}$ ($\boldsymbol{\kappa_{\max}})$} & \multicolumn{3}{|c|}{\textbf{Confidence intervals}}\\
		\cline{1-13}
		\textbf{Cardiologist} & 	
	\bigstrut 
	    C1 & C2  & C3 & C1 & C2  & C3 & 
	    C1 & C2  & C3 & C1 & C2  & C3\\ \hline
		\bigstrut\textbf{Morphology} & 0.25 (0.81) & 0.49 (0.79) & 0.73 (0.89) & (0.03, 0.46) & (0.28, 0.70) & (0.57, 0.89) & 0.80 (0.92) & 0.72 (0.87) & 0.76 (0.92) & (0.70, 0.91) & (0.60, 0.84) & (0.65, 0.86) \\ \hline
    \hline
	\bigstrut
		\multirow{2}{*}{\textbf{}} & \multicolumn{6}{|c||}{\textbf{T wave}} & \multicolumn{6}{|c|}{\textbf{ST segment}}\\
		\cline{2-13}
	\bigstrut
		& \multicolumn{3}{|c|}{$\boldsymbol{\kappa}$ ($\boldsymbol{\kappa_{\max}})$} & \multicolumn{3}{|c||}{\textbf{Confidence intervals}} & 		\multicolumn{3}{|c|}{$\boldsymbol{\kappa}$ ($\boldsymbol{\kappa_{\max}})$} & \multicolumn{3}{|c|}{\textbf{Confidence intervals}}\\
		\cline{1-13}
		\textbf{Cardiologist} & 	
	\bigstrut 
	    C1 & C2  & C3 & C1 & C2  & C3 & 
	    C1 & C2  & C3 & C1 & C2  & C3\\ \hline
		\bigstrut\textbf{Morphology} &  0.63 (0.86) & 0.69 (0.85) & 0.71 (0.87) & (0.47, 0.79) & (0.55, 0.84) & (0.56, 0.85) & 0.33 (0.76) & 0.37 (0.84) & 0.41 (0.78) & (0.04, 0.63) & (0.07, 0.67) & (0.16, 0.67) \\
		\bigstrut\textbf{Depressed} & - & - & - & - & - & - & 0.62 (0.87) & 0.77 (0.77) & 0.63 (0.63) & (0.42, 0.82) & (0.59, 0.95) & (0.34, 0.91) \\		\bigstrut\textbf{Elevated} & - & - & - & - & - & - & 0.43 (0.69) & 0.63 (0.93) & 0.67 (0.80) & (0.14, 0.71) & (0.32, 0.94) & (0.39, 0.95) \\		
		\hline
	\end{tabular}
}
\end{center}
\label{tab:between_method}
\vspace{-5mm}	
\end{table*}
\endgroup

\subsubsection{Inter-rater agreement}
\begin{figure}[!t]
\centerline{\includegraphics[width=\columnwidth, trim=90 320 110 320, clip]{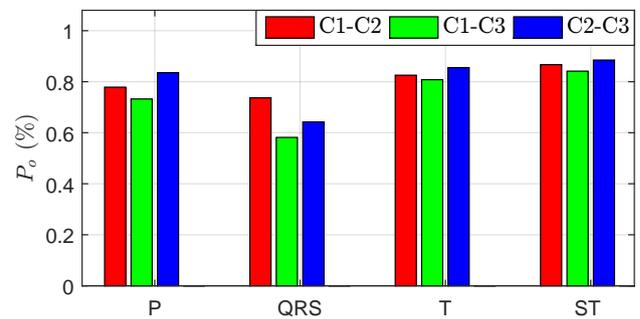}}
\caption{Inter-rater diagnostic concordance for the ECG features.}
\label{fig:between_observer_conc}
\end{figure}

We also evaluated the reproducibility of our tests by analyzing the inter-rater agreement. Fig.~\ref{fig:between_observer_conc} shows the inter-rater diagnostic concordance. As with the between-method concordance, the percent agreement is around $80\%$, except for the QRS complex, for which the value is considerably lower  than for the other features. The results show that the interpretation of these features can vary between cardiologists. For instance, a qRs-type QRS complex with very low-amplitude q and s waves can easily be misclassified as R, qR, or Rs wave depending on the stringency of the examiner. This explains the relatively low inter-rater concordance, which is also supported by the kappa statistics in Tab.~\ref{tab:between_obs}. The kappa values related to the QRS morphology are lower than those for the T wave, but the uncertainty of these estimates of $\kappa$ is also higher. Overall, in most cases there was moderate agreement between the cardiologists on the morphological features, except for the T wave, where the level of agreement was substantial. The discrepancies were probably caused by the different levels of medical experience and by the fact that the clinical standards and decision rules applied can vary between cardiologists. 

\begingroup
\setlength{\tabcolsep}{3pt} 
\renewcommand{\arraystretch}{1.5} 
\begin{table*}[t!]
\centering
\vspace{2mm}	
\caption{Using $\kappa$ statistics to analyze inter-rater agreement for the ECG features.}
\vspace{-2mm}
\begin{center}
\scalebox{0.73}{
	\begin{tabular}{|c|c|c|c|c|c|c||c|c|c|c|c|c|} \hline
	\bigstrut
		\multirow{2}{*}{\textbf{}} & \multicolumn{6}{|c||}{\textbf{P wave}} & \multicolumn{6}{|c|}{\textbf{QRS complex}}\\
		\cline{2-13}
	\bigstrut
		& \multicolumn{3}{|c|}{$\boldsymbol{\kappa}$ ($\boldsymbol{\kappa_{\max}})$} & \multicolumn{3}{|c||}{\textbf{Confidence intervals}} & 		\multicolumn{3}{|c|}{$\boldsymbol{\kappa}$ ($\boldsymbol{\kappa_{\max}})$} & \multicolumn{3}{|c|}{\textbf{Confidence intervals}}\\
		\cline{1-13}
		\textbf{Cardiologist} & 	
	\bigstrut 
	    C1-C2 & C1-C3 & C2-C3 & C1-C2 & C1-C3 & C2-C3 & 
	    C1-C2 & C1-C3 & C2-C3 & C1-C2 & C1-C3 & C2-C3\\ \hline
		\bigstrut\textbf{Morphology} & 0.53 (0.68) & 0.44 (0.79) & 0.65 (0.84) & (0.39, 0.68) & (0.28, 0.59) & (0.52, 0.77) & 0.67 (0.83) & 0.51 (0.63) & 0.58 (0.70) & (0.59, 0.76) & (0.42, 0.60) & (0.50, 0.67) \\ \hline
    \hline
	\bigstrut
		\multirow{2}{*}{\textbf{}} & \multicolumn{6}{|c||}{\textbf{T wave}} & \multicolumn{6}{|c|}{\textbf{ST segment}}\\
		\cline{2-13}
	\bigstrut
		& \multicolumn{3}{|c|}{$\boldsymbol{\kappa}$ ($\boldsymbol{\kappa_{\max}})$} & \multicolumn{3}{|c||}{\textbf{Confidence intervals}} & 		\multicolumn{3}{|c|}{$\boldsymbol{\kappa}$ ($\boldsymbol{\kappa_{\max}})$} & \multicolumn{3}{|c|}{\textbf{Confidence intervals}}\\
		\cline{1-13}
		\textbf{Cardiologist} & 	
	\bigstrut 
	    C1-C2 & C1-C3 & C2-C3 & C1-C2 & C1-C3 & C2-C3 & 
	    C1-C2 & C1-C3 & C2-C3 & C1-C2 & C1-C3 & C2-C3 \\ \hline
		\bigstrut\textbf{Morphology} &  0.71 (0.95) & 0.67 (0.92) & 0.76 (0.88) & (0.60, 0.81) & (0.56, 0.78) & (0.66, 0.85) & 0.48 (0.81) & 0.42 (0.82) & 0.49 (0.59) & (0.27, 0.68) & (0.23, 0.60) & (0.31, 0.68) \\
		\bigstrut\textbf{Depressed} & - & - & - & - & - & - & 0.69 (0.79) & 0.46 (0.50) & 0.58 (0.67) & (0.55, 0.82) & (0.28, 0.64) & (0.39, 0.76) \\		\bigstrut\textbf{Elevated} & - & - & - & - & - & - & 0.57 (0.67) & 0.69 (0.74) & 0.72 (0.93) & (0.36, 0.77) & (0.51, 0.86) & (0.53, 0.91) \\		
		\hline
	\end{tabular}
}
\end{center}
\label{tab:between_obs}
\vspace{-5mm}	
\end{table*}
\endgroup

\subsubsection{Within-observer agreement}
\begin{figure}[!t]
\centerline{\includegraphics[width=\columnwidth, trim=90 320 110 320, clip]{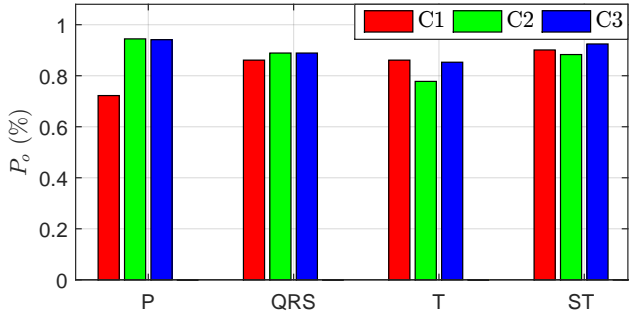}}
\caption{Within-observer diagnostic concordance for ECG features.}
\label{fig:within_observer_conc}
\end{figure}

To assess the repeatability of our test, we studied the within-observer concordance (Fig.~\ref{fig:within_observer_conc}) by using 12 repeated records including three leads. The agreement observed was much higher than in the between-method and inter-rater cases. This was to be expected, since an individual cardiologist's interpretation of ECG features should not vary much. The results indicate high self-consistency among the medical experts, which is also supported by the kappa statistics. In fact, $\kappa$ is very close to $\kappa_{\max}$ and suggests almost perfect agreement for the P wave and the QRS complex, and substantial agreement for the T wave and the ST segment. The kappa values are very low for C2 and C3 in the cases of general ST morphology and ST elevation, respectively. This is due to the first paradox of $\kappa$, which is caused by the high prevalence of the corresponding ST features. For instance, C3 considered ST elevation to be absent in almost all the repeated records, and agreed with himself in 35 of the overall 36 cases including the three leads. Although we would expect almost perfect agreement, the high prevalence reduces the value of kappa, since $P_c\approx P_o$ (see e.g.,~\cite{kappa_para1, kappa_para2}). For the same reason, $\kappa$ is not applicable (n/a) in the case of elevated ST for C2, who achieved perfect agreement with himself on the absence of elevated ST in 36 out of 36 cases. This implies perfect agreement, but the denominator in Eq.~\eqref{eq:kappa} becomes zero due to $P_o=1$. In summary, we observed a high level of self-consistency among the cardiologists, which demonstrates the robustness of this study.

\begingroup
\setlength{\tabcolsep}{3pt} 
\renewcommand{\arraystretch}{1.5} 
\begin{table*}[t!]
\centering
\vspace{2mm}	
\caption{Using $\kappa$ statistics to analyze within-observer agreement for the ECG features.}
\vspace{-2mm}
\begin{center}
\scalebox{0.73}{
	\begin{tabular}{|c|c|c|c|c|c|c||c|c|c|c|c|c|} \hline
	\bigstrut
		\multirow{2}{*}{\textbf{}} & \multicolumn{6}{|c||}{\textbf{P wave}} & \multicolumn{6}{|c|}{\textbf{QRS complex}}\\
		\cline{2-13}
	\bigstrut
		& \multicolumn{3}{|c|}{$\boldsymbol{\kappa}$ ($\boldsymbol{\kappa_{\max}})$} & \multicolumn{3}{|c||}{\textbf{Confidence intervals}} & 		\multicolumn{3}{|c|}{$\boldsymbol{\kappa}$ ($\boldsymbol{\kappa_{\max}})$} & \multicolumn{3}{|c|}{\textbf{Confidence intervals}}\\
		\cline{1-13}
		\textbf{Cardiologist} & 	
	\bigstrut 
	    C1 & C2  & C3 & C1 & C2  & C3 & 
	    C1 & C2  & C3 & C1 & C2  & C3\\ \hline
		\bigstrut\textbf{Morphology} & 0.49 (0.49) & 0.82 (0.91) & 0.82 (0.82) & (0.22, 0.76) & (0.58, 1.00) & (0.58, 1.00) & 0.82 (0.89) & 0.85 (0.89) & 0.86 (0.86) & (0.67, 0.97) & (0.72, 0.99) & (0.74, 0.99) \\ \hline
    \hline
	\bigstrut
		\multirow{2}{*}{\textbf{}} & \multicolumn{6}{|c||}{\textbf{T wave}} & \multicolumn{6}{|c|}{\textbf{ST segment}}\\
		\cline{2-13}
	\bigstrut
		& \multicolumn{3}{|c|}{$\boldsymbol{\kappa}$ ($\boldsymbol{\kappa_{\max}})$} & \multicolumn{3}{|c||}{\textbf{Confidence intervals}} & 		\multicolumn{3}{|c|}{$\boldsymbol{\kappa}$ ($\boldsymbol{\kappa_{\max}})$} & \multicolumn{3}{|c|}{\textbf{Confidence intervals}}\\
		\cline{1-13}
		\textbf{Cardiologist} & 	
	\bigstrut 
	    C1 & C2  & C3 & C1 & C2  & C3 & 
	    C1 & C2  & C3 & C1 & C2  & C3\\ \hline
		\bigstrut\textbf{Morphology} &  0.79 (0.92) & 0.67 (0.79) & 0.78 (0.82) & (0.62, 0.96) & (0.46, 0.87) & (0.60, 0.96) &  0.73 (0.91) & 0.08 (0.28) & 0.78 (0.78) & (0.44, 1.00) & (-0.38, 0.54) & (0.54, 1.00) \\
		\bigstrut\textbf{Depressed} & - & - & - & - & - & - & 0.66 (1.00) & 0.88 (1.00) & 0.61 (0.61) & (0.42, 0.91) & (0.72, 1.00) & (0.25, 0.97) \\		\bigstrut\textbf{Elevated} & - & - & - & - & - & - & 0.79 (0.79) & n/a (n/a) & 0.00 (0.00) & (0.37, 1.00) & (n/a, n/a) & (-1.00, 1.00) \\		
		\hline
	\end{tabular}
}
\end{center}
\label{tab:within_obs}
\vspace{-5mm}	
\end{table*}
\endgroup

\section{Discussion \label{sec:discussions}}
Although $\kappa$ is the most commonly used agreement measure in the literature, it is often criticized as being somewhat difficult to interpret in particular situations~\cite{kappa_studies1}. For instance, the prevalence of the attributes affects the magnitude of $\kappa$, which was in fact the case with the ST morphologies in Tab.~\ref{tab:within_obs}. This effect becomes apparent when the proportion
of agreements on one attribute differs significantly from those for the others. High prevalence causes high chance
agreement $P_c$, which reduces the value of $\kappa$ accordingly. Regarding our experiments on the various types of concordance, we found that the attributes of positive P and T waves and the horizontal ST segment had the highest prevalence. This was to be expected, since these are the most common waveforms in the ECG. In these cases, low values of $\kappa$ do not necessarily imply low rates of overall agreement.  Therefore, alongside the value of $\kappa$, we also reported the diagnostic concordance for each experiment in Figs.~\ref{fig:between_method_conc}-\ref{fig:within_observer_conc}. 
Note that there is greater potential for disagreement in the case of a large number of optional categories, as is the case for the QRS complex. Thus, the high values of $\kappa$ indicate very strong agreement especially for the QRS morphologies, where we used 10 different shape categories. Generally, in order to counteract high prevalence and the resulting low $\kappa$ values, this approach could be extended with interesting/rare \ac{ecg} recordings, which would result in a more heterogeneous set of wave shapes. 

\begin{figure}[!t]
\centerline{\includegraphics[width=\columnwidth, trim=90 300 110 320, clip]{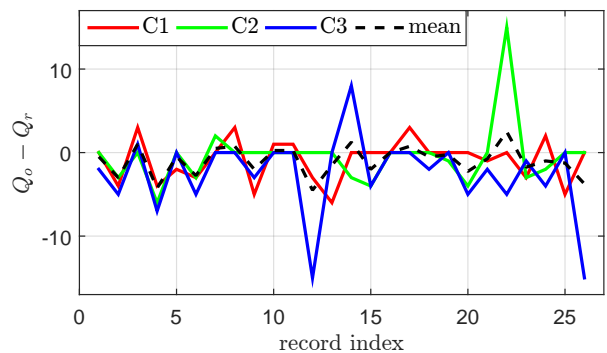}}
\caption{Quality score differences between the original and the filtered ECG signals.}
\label{fig:quality_score_diff}
\end{figure}

We also evaluated the general quality scores for the original signal ($Q_o$) and for the reconstructed ECG ($Q_r$). Fig.~\ref{fig:quality_score_diff} plots the differences $Q_o-Q_r$ for each cardiologist and for the mean. Excluding the outliers, the differences have negative signs, which indicates an improved quality in the case of the reconstructed signal \cite{carlpeter}. Although this visual enhancement is expected, filtering does not necessarily result in better diagnostic quality. For instance, FIR and IIR filters can remove the noise in the targeted frequency band, but they may also introduce ringing artifacts due to the well-known Gibbs phenomenon \cite{ecgbook}. However, the between-method agreement study and the quality score differences show that our joint Hermite-sigmoid model~\cite{carlpeter} represents the ECG in a low-dimensional space without diagnostic distortion, and even enhances the visual quality of the ECG in most cases. 

\begin{figure}[!t]
\centerline{\includegraphics[width=\columnwidth]{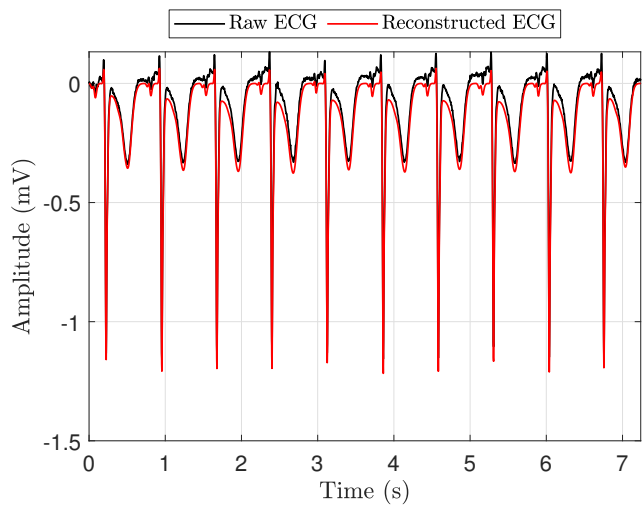}}
\caption{For (very) low and high frequent noise (not in the relevant ECG subbands) the wavelet-based objective measures accurately capture the low diagnostic distortion of the low-dimensional ECG representation (WWPRD=$9.46\%$, WEDD=$7.89\%$), while PRD=$14.38\%$ already indicates a diagnostic distortion.}
\label{fig:mgh056_cmp_related_work}
\end{figure}

\begin{figure}[!t]
\centerline{\includegraphics[width=\columnwidth]{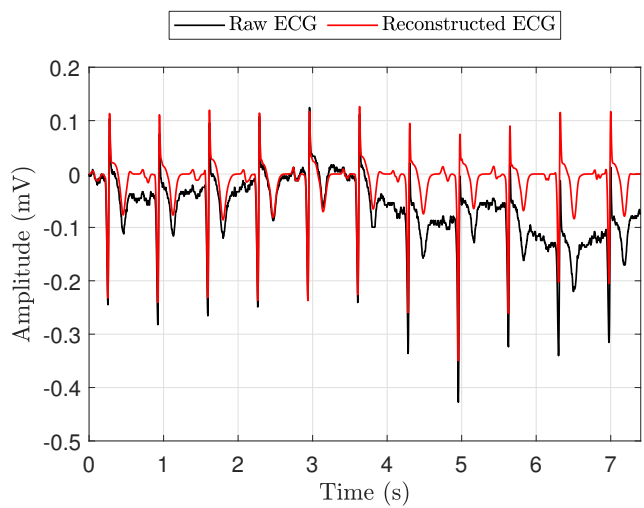}}
\caption{The high values of the objective measures (PRD=$67.5\%$, WWPRD=$29.0\%$, WEDD=$59.7\%$) indicate a high diagnostic distortion, although the low dimensional representation perfectly removes the noise and keeps the most important diagnostic features which is in line with the results of our framework.}
\label{fig:mgh184_cmp_related_work}
\end{figure}

Hence, compared to the objective measures  PRD, WWPRD \cite{wwprd}, and WEDD \cite{wedd}, we obtain a more reliable assessment of the diagnostic distortion and signal quality of the reconstructed signal. While Fig.~\ref{fig:mgh056_cmp_related_work} illustrates that specifically the wavelet-based objective measures perform well in case of (very) low and very high frequent noise (record index 8, mgh056), which is not overlapping with the relevant \ac{ecg} subbands, Fig.~\ref{fig:mgh184_cmp_related_work} reveals their weaknesses (record index 20, mgh184). In this case the ECG is superimposed by baseline wander noise that overlaps with relevant \ac{ecg} subbands, therefore the objective measures show a high diagnostic distortion. In fact, the high values (PRD=$67.5\%$, WWPRD=$29.0\%$, WEDD=$59.7\%$) all correspond to low quality groups according to Tab.~8 in \cite{wedd}. However, the signal quality is significantly improved by the low dimensional representation (Fig.~\ref{fig:mgh184_cmp_related_work}), hence, the objective measures are misleading for this recording. In fact, the visual inspection of the original and the filtered ECGs conducted by three cardiologists confirmed the quality improvement for this record (see Fig.~\ref{fig:quality_score_diff}), which did not indicate any important diagnostic loss. Clearly, the objective measures are helpful and reliable in many cases (e.g., Fig.~\ref{fig:mgh056_cmp_related_work}), nevertheless, they are limited for noise overlapping with relevant \ac{ecg} subbands, demanding frameworks as suggested in this work that overcome this limitation.

Furthermore, as illustrated in Figure~\ref{fig:mos_test_example}, the questionnaire also offers the option of proposing a main diagnosis for the three \ac{ecg} leads. This is intended to be answered by medical experts only, who are briefed that they do not have to give a definitive answer, but their best guess based on the three leads. This provides an additional source for identifying diagnostic distortions which might not be covered by the standard questionnaire (in case diagnosis differs significantly between raw and reconstructed signals). Clearly, this question is challenging and should therefore -- if it is to bring value -- only be answered in the final stage of algorithm evaluation by experienced, meticulous experts. In our case, the three experts gave a main diagnosis, showing good agreement in most of the cases, that is, $73\,\%$, $88\,\%$ and $92\,\%$, respectively. A more detailed analysis by a medical expert, who subsequently investigated why the raw and reconstructed \ac{ecg} recordings led to different diagnoses uncovered a loss of diagnostic information in the ST segment. On this basis our previously published algorithm \cite{carlpeter} can be further improved in the future. Alesanco et al. \cite{mostest} argued that this third question should be presented in a semi-blind way, which means, showing both the raw and the reconstructed \ac{ecg}s to the medical expert at the same time and asking whether they would evaluate them differently. This could potentially reduce intra-subject variability caused, for instance, by varying levels of attention when judging the raw and the reconstructed recordings. However, the drawback of this is that -- even unintentionally -- one typically tends to find the same characteristics in the two recordings if they are presented at the same time. Consequently, this may lead to misjudging an important diagnostic loss, and therefore we believe that the blind test is more appropriate in this case.

\section{Conclusion \label{sec:conc}}
We have proposed a methodology for 
quantifying the diagnostic distortion 
of ECG signal processing algorithms, for instance, for filtering, segmentation, and data compression. These low-dimensional ECG representations may lead to distortion of the diagnostic information contained in the ECG, which we quantified using Cohen's kappa. Note that $\kappa$ is affected by prevalence, and thus the corresponding quality scores are not meant to perform direct comparisons between ECG processing algorithms of different studies.    
Instead, our goal was to design a testing framework that includes a questionnaire which minimizes the time to train non-medical staff to evaluate the diagnostic distortion of ECG decompositions. To this end, we chose scoring rubrics such that the interpretation of original and reconstructed ECGs is clear and some level of objectivity is imposed on the rating scale. The proposed test is therefore free of ambiguous medical features, such as symmetry and notches. In a case study, we considered Hermite-based characterization of ECG waveforms, which is a very popular topic in this field. Particularly, we showed that low-dimensional heartbeat signal representation by means of Hermite and sigmoidal functions preserves diagnostically relevant features of the ECG.

\bibliographystyle{model1-num-names}

\bibliography{refs}

\end{document}